\documentclass[conference]{IEEEtran}
\IEEEoverridecommandlockouts
\usepackage{cite}
\usepackage{geometry}
\usepackage{setspace}
\usepackage{amsmath,amssymb,amsfonts}
\usepackage{algorithmic}
\usepackage{graphicx}
\usepackage{textcomp}
\usepackage{xcolor}
\usepackage{amsthm}
\usepackage{caption}
\usepackage{float} 
\usepackage{subcaption}

\newtheorem{lemma}{Lemma}
\newtheorem{theorem}{Theorem}

\newcounter{MYtempeqncnt}
\def\BibTeX{{\rm B\kern-.05em{\sc i\kern-.025em b}\kern-.08em
		T\kern-.1667em\lower.7ex\hbox{E}\kern-.125emX}}

\begin{document}
	
	\title{Performance Analysis of Integrated Data and Energy Transfer Assisted by Fluid Antenna Systems
	}
	
	\author{\IEEEauthorblockN{Xiao Lin$^1$, Halvin Yang$^{2}$, Yizhe Zhao$^1$, Jie Hu$^{1,*}$, Kai-Kit Wong$^{2}$ \thanks{This work was supported in part by the Key Research and Development Program of Zhejiang Province under Grant 2022C01093, in part by the Natural Science Foundation of China under Grant 62132004 and Grant 62201123, in part by Young Elite Scientists Sponsorship Program by CAST under Grant 2023QNRC001, in part by the Engineering and Physical Sciences Research Council (EPSRC) under Grant EP/W026813/1, in part by the Stable Supporting Fund of National Key Laboratory of Underwater Acoustic Technology; in part by China Postdoctoral Science Foundation under Grant 2022TQ0056.}} \\
		\IEEEauthorblockA{$^1$\small School of Information and Communication Engineering, University of Electronic Science and Technology of China (UESTC), \\
			Chengdu, Sichuan, China \\
			$^{2}$Department of Electronic and Electrical Engineering, University College London (UCL), London, United Kingdom  \\
			$^{*}$Corresponding Author, Email: hujie@uestc.edu.cn
		}
	}
	\maketitle
	\begin{abstract}
		Fluid antenna multiple access (FAMA) is capable of exploiting the high spatial diversity of wireless channels to mitigate multi-user interference via flexible port switching, which achieves a better performance than traditional multi-input-multi-output (MIMO) systems. Moreover, integrated data and energy transfer (IDET) is able to provide both the wireless data transfer (WDT) and wireless energy transfer (WET) services towards low-power devices. In this paper, a FAMA assisted IDET system is studied, where $N$ access points (APs) provide dedicated IDET services towards $N$ user equipments (UEs). Each UE is equipped with a single fluid antenna. The performance of WDT and WET , \textit{i.e.}, the WDT outage probability, the WET outage probability, the reliable throughput and the average energy harvesting amount, are analysed theoretically by using time switching (TS) between WDT and WET. Numerical results validate our theoretical analysis, which reveals that the number of UEs and TS ratio should be optimized to achieve a trade-off between the WDT and WET performance. Moreover, FAMA assisted IDET achieves a better performance in terms of both WDT and WET than traditional MIMO with the same antenna size.
	\end{abstract}
	
	
	\section{Introduction}
	In the era of 6G, massive low-power devices are swarming into the networks for providing various services to smart cities, posing challenges to network spectrum efficiency and energy efficiency \cite{Z.Zhang,W.Saad}. Traditional multi-input-multi-output (MIMO) techniques can achieve spatial multiplexing and spatial diversity via pre-coding at the transceiver, which improves the communication performance of the crowded network. However, multi-antennas are required at the receiver to achieve the considerable network gain, which is impractical for the low-power devices having limited hardware size.
	
	Fluid antenna is considered as a new promising technology to tackle with this hard problem, where any software-controllable fluidic and conductive radiative structure is able to change its position to reconfigure the radiation pattern. These potential switching positions are defined as ports. It is demonstrated that a single small size of fluid antenna can outperform  maximum ratio combining (MRC) assisted MIMO system when the number of ports is large enough \cite{FAsystem}.
	Recently, fluid antenna multiple access (FAMA) has emerged as an innovative technique to meet the requirements of the future crowded network, where the receiver is capable of selecting the optimal ports to mitigate the interference and enhance the strength of its own receive signals. Wong \textit{et al.} studied the system where the fluid antenna at each user is switched to an optimal port on a symbol-by-symbol basis, namely fast-FAMA ($f$-FAMA) \cite{FAMA}. By contrast, recently, Wong \textit{et al.} proposed a new FAMA system, namely slow-FAMA ($s$-FAMA), where each user switches its fluid antenna port only if the envelopes of the wireless channel changes \cite{gkbykit}. It was demonstrated that by conceiving sufficient resolution and size of the fluid antenna, FAMA could support a massive number of users by using only a single fluid antenna. In paper \cite{Massive}, the authors proposed a simple approach to estimate the instantaneous sum-interference plus noise signals for port selection at every symbol instance to unleash the performance of $f$-FAMA. Yang \textit{et al.} derived new closed-form expressions for the outage probability with noise consideration in $s$-FAMA \cite{yang}. Futhermore, Wong \textit{et al.} quantified the benefits of the synergy between opportunistic scheduling and FAMA by analyzing the multiplexing gain of the network \cite{Oppo}.
	
   	Nowadays, the energy consumption of low-power devices poses a significant challenge.  Several papers have proposed diverse methods to enhance performance while minimizing energy consumption \cite{wu1,wu2}. The energy supplement is a straightforwrd approach to address  the issue of insufficient energy. Nevertheless, energy supplement of all the massive low-power devices is impossible in the future crowded network, since artificially charging or replacing batteries requires undeserved human costs.
	
	Wireless energy transfer (WET) is considered as a promising solution to realize convenient and wide-area power supply for the low-power devices. Since they have both data and energy requirements, wireless data transfer (WDT) and WET should be coordinated at the transceiver, which yields the concept of integrated data and energy transfer (IDET) \cite{Rzhang}. Zong \emph{et al.} investigated the transceiver design problem for IDET in $K$-user MIMO interference channels \cite{Optimal}. Hu \emph{et al.} provided the first detailed survey on the key techniques of IDET communication networks \cite{hujie}. However, the receive signals suffer from the serious power attenuation in the transmission path, which degrade the IDET efficiency, especially for WET whose activating power threshold is much higher than that of WDT. FAMA is able to tune
	in to the window of opportunity in which the interference naturally disappears in a deep fade and enhance the receive signal strength by flexibly adjusting the port selection, which improves the IDET performace in the multi-user scenario.
	
	 In this paper, we propose a FAMA assisted multi-user IDET system. Time switching (TS) approach is applied for separating WET and WDT, where each transmission block is divided into two time slots, one for WDT and  the other for WET. During each time slot, the fluid antenna of each UE independently switches to an optimal port for either WDT or WET. Then, our novel contributions are then summarized as follows:	

	\begin{itemize}
		\item[$\bullet$] We propose a FAMA assisted multi-user IDET system, where TS approach is conceived for separating WDT and WET to improve the system performance.
		\item[$\bullet$] The IDET performance, \textit{i.e.}, the WDT outage probability, the WET outage probability, the reliable throughput and the average energy harvesting amount, are all analyzed theoretically into the approximated closed-forms.
		\item[$\bullet$] The theoretical analysis is validated by the Monte Carlo based simulation, while the IDET performance of the FAMA assisted IDET system is also evaluated.
	\end{itemize}
	\section{System Model}
\begin{figure}[t]
		\centering
		\includegraphics[width=3.1in]{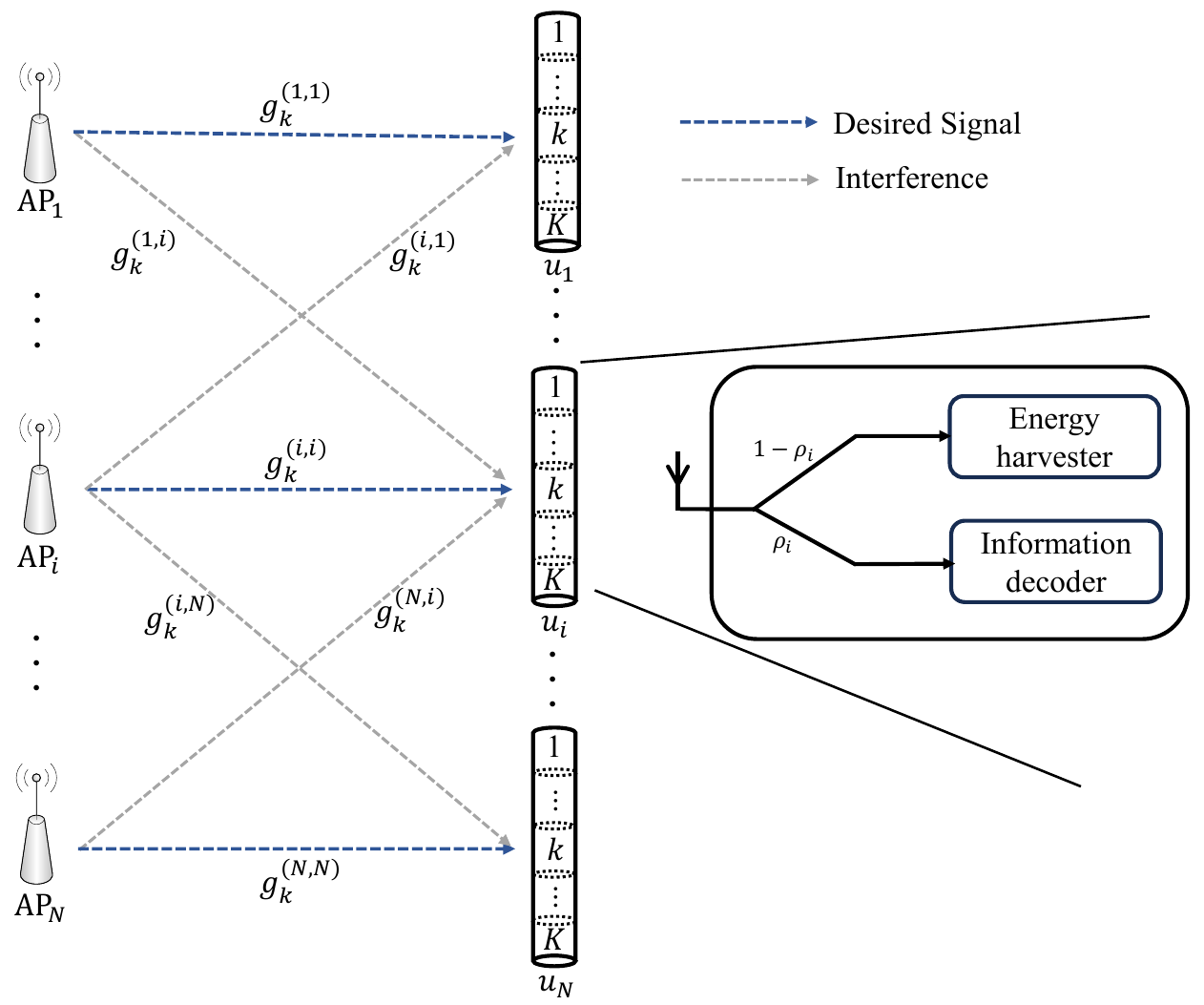}
		\caption{System model of FAMA assisted IDET in the multi-user scenario.}
		\label{model}
	\end{figure}

	We study a downlink FAMA assisted IDET system consists of $N$ fixed access points (APs) and $N$ user equipments (UEs). Each UE is equipped with a $K$-port linear fluid antenna having the size of $W\lambda$, where $W$ denotes the normalized size of the fluid antenna and $\lambda$ denotes the wavelength. It is assumed that each AP is dedicated to communicate with its corresponding UE, \textit{i.e.}, $\text{AP}_i$ communicates with $u_i$, and each UE applies TS approach to coordinate WDT and WET, as shown in Fig. \ref{model}. The time duration of WDT and WET of $u_i$ is denoted as $\alpha_iT$ and $(1-\alpha_i)T$ respectively, where $\alpha_i$ denotes the TS ratio and $T$ is the duration of the whole period. UEs may also receive the interference signals from other APs, which may degrade the WDT performance but provide more signal sources for WET.
	\subsection{Wireless channel model}
	The additional white Gaussian noise (AWGN) channel is conceived in our system\footnote{In this paper, we focus on the impact of channel fading on the IDET performance, while that of pathloss is neglected.}, while the wireless channel gain  between $\text{AP}_m$ and the $k$-th port of $u_i$ is characterized by \cite{gkbykit} as
	\begin{equation}\label{eqgk}
		\begin{aligned}
			g_k^{(m,i)}=(\sqrt{1-\mu^2}&x_k^{(m,i)}+\mu x_0^{(m,i)})\\
			+j(&\sqrt{1-\mu^2}y_k^{(m,i)}+\mu y_0^{(m,i)}),
		\end{aligned}
	\end{equation}
	where $x_0^{(m,i)},\cdots,x_K^{(m,i)}$ and $y_0^{(m,i)},\cdots,y_K^{(m,i)}$ are all independent Gasussian random variables having zero mean and the variance of $1$. Note that the correlation among different ports is realized by invoking the same variables $x_0^{(m,i)} , y_0^{(m,i)}$. $\mu$ represents the correlation parameter among different ports, which is expressed by \cite{mu} as
	\begin{equation}
		\begin{aligned}
			\mu=\sqrt{2}\sqrt{_{1}F_2(\frac{1}{2};1,\frac{3}{2};-\pi^2W^2)-\frac{J_1(2\pi W)}{2\pi W}},
		\end{aligned}
	\end{equation}
	where $_{a}F_b(\cdot;\cdot;\cdot)$ denotes the generalized hypergeometric function and $J_1(\cdot)$ is the first-order Bessel function of the first kind.
	\subsection{IDET with TS approach}	
 TS approach is conceived at each UE to separate WET and WDT in a transmission period. Then, in each period, the received signal at $u_i$ by activating the $k$-th port is expressed as
	\begin{equation}
		\begin{aligned}
			y_{k,i}=\underbrace{\sqrt{P_i}g_k^{(i,i)}s_i}_{\text{desired signal by AP}_i}+\underbrace{\sum_{m\neq i}\sqrt{P_m}g_k^{(m,i)}s_m}_{\text{interference caused by the rest APs}}+\underbrace{\eta_k^i+z_i}_{\text{noise}},
		\end{aligned}
	\end{equation}
	where $s_i$ is the transmitted Gaussian signal having the unit power for $u_i$, $P_i$ denotes the transmission power of $s_i$. $\eta_k^i$ is the complex AWGN having the zero mean and the variance of $\sigma^2_\eta$, at the $k$-th antenna port of $u_i$. $z_i$ is the passband-to-baseband noise of $u_i$ which also follows the complex Gaussian distribution having the zero mean and the variance of $\sigma^2_c$.
	
	In each period, information decoding is firstly operated at the WDT slot of $u_i$. If the $k$-th antenna port is selected, the signal-to-interference-plus-noise-ratio (SINR) of $u_i$  is expressed as
	\begin{equation}\label{SINR}
		\begin{aligned}
			\text{SINR}_k^i&=\frac{P_i \vert g_k^{(i,i)} \vert^2 \mathbb{E}[\Vert s_i \Vert ^2]}{\sum_{m\neq i}^{N}P_m\vert g_k^{(m,i)} \vert^2 \mathbb{E}[\Vert s_m \Vert ^2]+\sigma_{\eta}^2+\sigma_{c}^2}\\
			&\overset{(a)}{\approx}\frac{P_i \vert g_k^{(i,i)} \vert^2 }{\sum_{m\neq i}^{N}P_m\vert g_k^{(m,i)} \vert^2 },
		\end{aligned}
	\end{equation}
	where $(a)$ assumes that the interference power is much greater than the noise power, from which the SINR becomes the signal-to-interference-ratio (SIR). The UE $u_i$ then operates energy harvesting in the next WET slot. If the $k$-th antenna port is selected, the energy harvesting power is then expressed as
	\begin{equation}\label{harvestpower}
		\begin{aligned}
			Q_k^i=\sum_{m=1}^{N}P_m\vert g_k^{(m,i)} \vert^2.
		\end{aligned}
	\end{equation}
	\subsection{Port slection}
	Interference signals may result in different impact on the performance of WDT and WET. For instance, a stronger interference may cause information decoding failure, while it may increase the receive power for WET. Therefore, port selection method of the fluid antenna should be separately designed for either WDT or WET. In the WDT slot, $u_i$ chooses an optimal port that maximizes the SINR expression in \eqref{SINR} as
	\begin{equation}
		\begin{aligned}
			k_{WDT}^{\ast}&= \arg\max_k \enspace \frac{P_i \vert g_k^{(i,i)} \vert^2 }{\sum_{m\neq i}^{N}P_m \vert g_k^{(m,i)} \vert^2 }\\
			&\overset{(b)}{=} \arg\max_k \enspace \frac{\vert g_k^{(i,i)} \vert^2 }{\sum_{m\neq i}^{N} \vert g_k^{(m,i)} \vert^2},\\
		\end{aligned}
	\end{equation}
	where $(b)$ assumes the case that all the APs have identical transmission power $P$. In the WET slot, $u_i$ selects an optimal antenna port that maximizes the total energy harvesting power in \eqref{harvestpower} as
	\begin{equation}	
		\begin{aligned}
			k_{WET}^{\ast}= \arg\max_k \enspace P\sum_{m=1}^{N}\vert g_k^{(m,i)} \vert^2.
		\end{aligned}
	\end{equation}
	\section{Performance Analysis of the FAMA Assisted IDET System}
	\subsection{WDT Performance Analysis}
	To ensure the correctness of information decoding, the SINR at each UE should be as high as possible. The WDT outage probability $\epsilon^i_{WDT}$ is defined as the probability that the SINR at $u_i$ is lower than a threshold $\gamma$, which is expressed as
	\begin{equation}
		\begin{aligned}
			\epsilon^i_{WDT}&=\text{Prob}(\text{SINR}^i_{k^*_{WDT}}<\gamma)\\
			&=\text{Prob}(\max_{k} \enspace \text{SINR}_k^i < \gamma)\\
			&=\text{Prob}(\max_{k} \enspace \frac{\vert g_k^{(i,i)} \vert^2}{\sum_{m\neq i}^{N}\vert g_k^{(m,i)} \vert^2} < \gamma).\\
		\end{aligned}
	\end{equation}
	According to \cite{gkbykit}, the upper bound of $\epsilon^i_{WDT}$ is derived as
	\begin{equation}\label{WDTP}
		\begin{aligned}
			\epsilon^i_{WDT}=[1-K(\frac{\mu^2}{\gamma+1})^{N-1}-K(\frac{1-\mu^2}{\gamma})^{N-1}]^{+},
		\end{aligned}
	\end{equation}
	where $[c]^{+}=\text{max}\{0,c\}$. Then, we aim to analyse the reliable throughput of the FAMA-IDET system, defined as $\tau=N\epsilon^{out}_{WDT}\log(1+\gamma)$, where $\gamma$ is the SINR threshold and $\epsilon^{out}_{WDT}$ represents the probability that at least one UE suffers from WDT outage \cite{Relaying}. Accordingly, $\epsilon^{out}_{WDT}$ can be expressed as
	\begin{equation}\label{throughput1}
		\begin{aligned}
			\epsilon^{out}_{WDT}&=1-(1-\epsilon^1_{WDT})(1-\epsilon^2_{WDT})\cdots(1-\epsilon^N_{WDT})\\
			&=1-\{\min[1,K(\frac{\mu^2}{\gamma+1})^{N-1}+K(\frac{1-\mu^2}{\gamma})^{N-1}]\}^N.
		\end{aligned}
	\end{equation}
	Given that only a fraction of $\alpha_i$ is allocated for WDT of each UE, the reliable throughput in the whole period is then formulated as
	\begin{equation}\label{throughput2}
		\begin{aligned}
			\tau&=NR\alpha(1-\epsilon^{out}_{WDT})\\
			&=NR\alpha\{\min[1,K(\frac{\mu^2}{\gamma+1})^{N-1}+K(\frac{1-\mu^2}{\gamma})^{N-1}]\}^N,
		\end{aligned}
	\end{equation}
	where the TS ratios are assumed to be the same as, $\alpha_i=\alpha, (i=1, 2, \cdots, N)$  for the sake of simplicity.
	\subsection{WET Performance Analysis}
	In the WET slot, the UE harvests the power of the received signals. However, due to hardware limitations, the power of the received signals should exceed a threshold, denoted as $Q_{th}$, in order to activate the energy harvesting circuits. Then, the WET outage probability is defined as the probability that the energy harvesting power $Q^i_{k^*_{WET}}$ is lower than the threshold $Q_{th}$. By conceiving the port selection method for WET, the energy harvesting of $u_i$ is futher formulated as
	\begin{equation}\label{eng2}
		\begin{aligned}
			Q^i_{k^*_{WET}}&=\max_{k}\enspace (1-\mu^2)P\sum_{m=1}^{N}[(x_k^{(m,i)}+\frac{\mu}{\sqrt{1-\mu^2}}x_0^{(m,i)})^2\\
			&\quad\quad\quad\quad\quad+(y_k^{(m,i)}+\frac{\mu}{\sqrt{1-\mu^2}}y_0^{(m,i)})^2].
		\end{aligned}
	\end{equation}
	By defining 
	\begin{equation}\label{exYk}
		\begin{aligned}
			Y_k=\sum_{m=1}^{N}[(&x_k^{(m,i)}+\frac{\mu}{\sqrt{1-\mu^2}}x_0^{(m,i)})^2\\
			+(&y_k^{(m,i)}+\frac{\mu}{\sqrt{1-\mu^2}} y_0^{(m,i)})^2],
		\end{aligned}
	\end{equation}
	the WET outage probability $\epsilon^i_{WET}$ of $u_i$ is then expressed as
	\begin{equation}\label{Qdef}
		\begin{aligned}
			\epsilon^i_{WET}=&\text{Prob}(Q^i_{k^*_{WET}}
			= \max_{k} \enspace (1-\mu^2)PY_k < Q_{th})\\
			=&\text{Prob}(\max_{k} \enspace Y_k < \frac{Q_{th}}{P(1-\mu^2)}=\overline{Q_{th}}).\\	
		\end{aligned}
	\end{equation}
	Note that given $x_0$ and $y_0$, $Y_k$ is non-central chi distributed with $2N$ degrees of freedom \cite[p. 22]{HGV}. By denoting $r_0\triangleq \sum_{m=1}^{N} (x_0^{(m,i)})^2+(y_0^{(m,i)})^2$, Then, the probability density function (PDF) of $Y_k$ is then formulated as
	\begin{equation}
		\begin{aligned}
			f_{Y_k\vert r_0}(h)=\frac{1}{2}(\frac{h}{\frac{\mu^2}{1-\mu^2}r_0})^{\frac{N-1}{2}}\exp{(-\frac{h+\frac{\mu^2}{1-\mu^2}r_0}{2})}\\
			\times I_{N-1}(\sqrt{\frac{\mu^2}{1-\mu^2}r_0h}),
		\end{aligned}
	\end{equation}
	where $I_{N-1}(\cdot)$ denotes the ($N-1$)-order modified Bessel function of the first kind. When $r_0$ is given and determined, all the $Y_k, (k=1,2\cdots,K)$ are independent with each other. Hence, the joint PDF of $Y_1, Y_2, \cdots, Y_K$ conditioned on $r_0$ is formulated as
	\begin{equation}\label{pdffk}
		\begin{aligned}
			f_{Y_1,\cdots,Y_K\vert r_0}&(h_1,\cdots,h_K)=\prod_{k=1}^{K}\frac{1}{2}(\frac{h_k}{\frac{\mu^2}{1-\mu^2}r_0})^{\frac{N-1}{2}}\times\\
			&\exp{(-\frac{h_k+\frac{\mu^2}{1-\mu^2}r_0}{2})}I_{N-1}(\sqrt{\frac{\mu^2}{1-\mu^2}r_0h_k}).
		\end{aligned}
	\end{equation}
	\begin{lemma}\label{lem1} 
		The conditional cumulative density function (CDF) of $Y_1,\cdots,Y_K$ is given by
		\begin{equation}
			\begin{aligned}\label{CDFFYk}
				F_{Y_1,\cdots,Y_K\vert r_0}&(t_1,\cdots,t_K)=\prod_{k=1}^{K}(1-Q_N(\sqrt{\frac{\mu^2}{1-\mu^2}r},\sqrt{t_k})),
			\end{aligned}
		\end{equation}
		where $Q_N(\cdot)$ is the $N$-order Marcum Q-function.	
	\end{lemma}
	Note that $r_0\triangleq \sum_{m=1}^{N} (x_0^{(m,i)})^2+(y_0^{(m,i)})^2$ is central chi distributed with $2N$ degrees of freedom \cite[p. 21]{HGV}. Therefore, we have
	\begin{equation}\label{pdfr0}
		\begin{aligned}
			f_{r_0}(r)=\frac{r^{N-1}\exp{(-\frac{r}{2})}}{2^N\Gamma(N)},
		\end{aligned}
	\end{equation}
	where $\Gamma(\cdot)$ is the gamma function. 
	\begin{theorem}\label{th1}
		The WET outage probability $\epsilon^i_{WET}$ of $u_i$ is formulated as
		\begin{equation}\label{final}
			\begin{aligned}
				\epsilon^i_{WET}=\int_{0}^{\infty}&\frac{r^{N-1}\exp{(-\frac{r}{2}})}{2^N\Gamma(N)}\times\\
				&[1-Q_N(\sqrt{\frac{\mu^2}{1-\mu^2}r},\sqrt{\overline{Q_{th}}})]^Kdr.
			\end{aligned}
		\end{equation}
	\end{theorem}
	
	\begin{IEEEproof}
		The WET outage probability is given by
		\begin{equation}
			\begin{aligned}\label{pfir}
				\epsilon^i_{WET}=\int_{0}^{\infty}f_{r_0}(r)F_{Y_1,\cdots,Y_K\vert r_0}&(\overline{Q_{th}},\cdots,\overline{Q_{th}})dr,
			\end{aligned}
		\end{equation}
		where $\overline{Q_{th}}=\frac{Q_{th}}{P(1-\mu^2)}$ is defined in \eqref{Qdef}.
		By substituting \eqref{pdfr0} and \eqref{CDFFYk} into \eqref{pfir} and after some simplification, \eqref{final} is obtained, which completes the proof.
	\end{IEEEproof}
	\begin{lemma}\label{lem2}
		In Gauss-Laguerre quadrature \cite[p. 923]{gaussian}.
		\begin{equation}\label{G1}
			\begin{aligned}
				\int_{0}^{\infty}g(x)dx\approx\sum_{l=1}^{n}w_l\exp{(\beta_l)}g(\beta_l),
			\end{aligned}
		\end{equation}
		where $\beta_l$ is the $l$-th root of Laguerre polynomial $L_n(x)$, and
		\begin{equation}\label{w1}
			\begin{aligned}
				w_l=\frac{\beta_l}{(n+1)^2(L_{n+1}(\beta_l))^2}.
			\end{aligned}
		\end{equation}
	\end{lemma}
	
	\begin{theorem}\label{th2}
		By invoking Lemma \ref{lem2}, the WET outage probability $\epsilon^i_{WET}$ can be approximated in a closed form as
		\begin{equation}\label{approp}
			\begin{aligned}
				\epsilon^i_{WET}=\sum_{l=1}^{n}w_l&\frac{\beta_l^{N-1}\exp{(\frac{\beta_l}{2}})}{2^N\Gamma(N)}\times \\
				&[1-Q_N(\sqrt{\frac{\mu^2}{1-\mu^2}\beta_l},\sqrt{\overline{Q_{th}}})]^K.
			\end{aligned}
		\end{equation}
	\end{theorem}	
	\begin{figure*}[!t]
		\normalsize
		\setcounter{MYtempeqncnt}{\value{equation}}
		\setcounter{equation}{23}
		\begin{small}
		\begin{equation}\label{formula: expectation}
			\begin{aligned}
				E_{avg}^i=\psi_i\int_{0}^{\infty}Kt&\int_{0}^{\infty}\frac{r^{N-1}\exp{(-\frac{r}{2}})}{2^{N+1}\Gamma(N)}[1-Q_N(\sqrt{\frac{\mu^2r}{1-\mu^2}},\sqrt{t})^{K-1}][Q_N(\sqrt{\frac{\mu^2r}{1-\mu^2}},\sqrt{t})-Q_{N-1}(\sqrt{\frac{\mu^2r}{1-\mu^2}},\sqrt{t})]drdt.\\
			\end{aligned}
		\end{equation}
		\end{small}
		\hrulefill
		\setcounter{equation}{27}
		\begin{small}
		\begin{equation}\label{formula: appro}
			\begin{aligned}
				E_{avg}^i\approx K\psi_i\sum_{u=1}^{n_2}\exp(\beta_u)w_u&\sum_{l=1}^{n_1}\frac{w_l\beta_l^{N-1}\exp{(\frac{\beta_l}{2}})}{2^{N+1}\Gamma(N)}[1-Q_N(\sqrt{\frac{\mu^2\beta_l}{1-\mu^2}},\sqrt{\beta_u})^{K-1}][Q_N(\sqrt{\frac{\mu^2\beta_l}{1-\mu^2}},\sqrt{\beta_u})-Q_{N-1}(\sqrt{\frac{\mu^2\beta_l}{1-\mu^2}},\sqrt{\beta_u})].\\
			\end{aligned}
		\end{equation}
		\end{small}
		\setcounter{equation}{\value{MYtempeqncnt}}
		\hrulefill
		\vspace*{4pt}
	\end{figure*}
	\subsection{Average energy harvesting amount}
	Then, we aim to analyse the average energy harvesting amount, $E_{avg}^i$, of $u_i$ in the WET slot, which is presented in Theorem \ref{th3}.
	\begin{theorem}\label{th3}
		The average energy harvesting amount, $E_{avg}^i$, of $u_i$  is expressed as \eqref{formula: expectation}.
	\end{theorem}
	\begin{IEEEproof}
		According to \eqref{eng2} and \eqref{exYk}, the average energy harvesting amount of $u_i$ is expressed as
		\setcounter{equation}{24}
		\begin{equation}
			\begin{aligned}
				E_{avg}^i&=(1-\alpha_i)T\mathbb{E}[Q_{k^*_{WET}}^i]\\
				&=(1-\alpha_i)(1-\mu^2)TP\mathbb{E}[\max_{k}\enspace \sum_{m=1}^{N}\vert g_k^{(m,i)} \vert^2]\\
				&=\psi_i\mathbb{E}[\max_{k}\enspace Y_k],\\
			\end{aligned}
		\end{equation}
		where $\psi_i=(1-\alpha_i)(1-\mu^2)TP$. By denoting  $Z=\max_k \enspace Y_k$, the CDF of variable $Z$ can be obtained in \eqref{final} by substituting $\overline{Q_{th}}$ with $z$. Thus,  the PDF of $Z$ is derived as
		\begin{equation}\label{deri}
			\begin{aligned}
				&f_Z(z)
				=\frac{\partial\int_{0}^{\infty}r^{N-1}\exp{(-\frac{r}{2}})[1-Q_N(\sqrt{\frac{\mu^2r}{1-\mu^2}},\sqrt{z})]^Kdr}{2^N\Gamma(N)\partial z}\\
				&=\int_{0}^{\infty}\frac{r^{N-1}\exp{(-\frac{r}{2}})}{2^N\Gamma(N)}\frac{\partial [1-Q_N(\sqrt{\frac{\mu^2r}{1-\mu^2}},\sqrt{z})]^K}{\partial z}dr\\
				&=K\int_{0}^{\infty}\frac{r^{N-1}\exp{(-\frac{r}{2}})}{2^{N+1}\Gamma(N)}[1-Q_N(\sqrt{\frac{\mu^2r}{1-\mu^2}},\sqrt{z})]^{K-1}]\\
				&\times[Q_N(\sqrt{\frac{\mu^2r}{1-\mu^2}},\sqrt{z})-Q_{N-1}(\sqrt{\frac{\mu^2r}{1-\mu^2}},\sqrt{z})]dr.
			\end{aligned}
		\end{equation}
		Then, $E_{avg}^i$ is formulated as \eqref{formula: expectation} by invoking
		\begin{equation}\label{simexp}
			\begin{aligned}
				\mathbb{E}[\max_{k}\enspace Y_k]=\mathbb{E}[Z]=\int_{0}^{\infty}zf(z)dz,\\
			\end{aligned}
		\end{equation}
		which completes the proof.
	\end{IEEEproof}
	
	\begin{theorem}\label{th4}
		By invoking Lemma \ref{lem2}, average energy harvesting amount of $u_i$ can be approximated in a closed form as \eqref{formula: appro}.
	\end{theorem}
	
	\section{Numerical Results}
	\begin{figure}[t]
		\centering
		\includegraphics[width=2.45in]{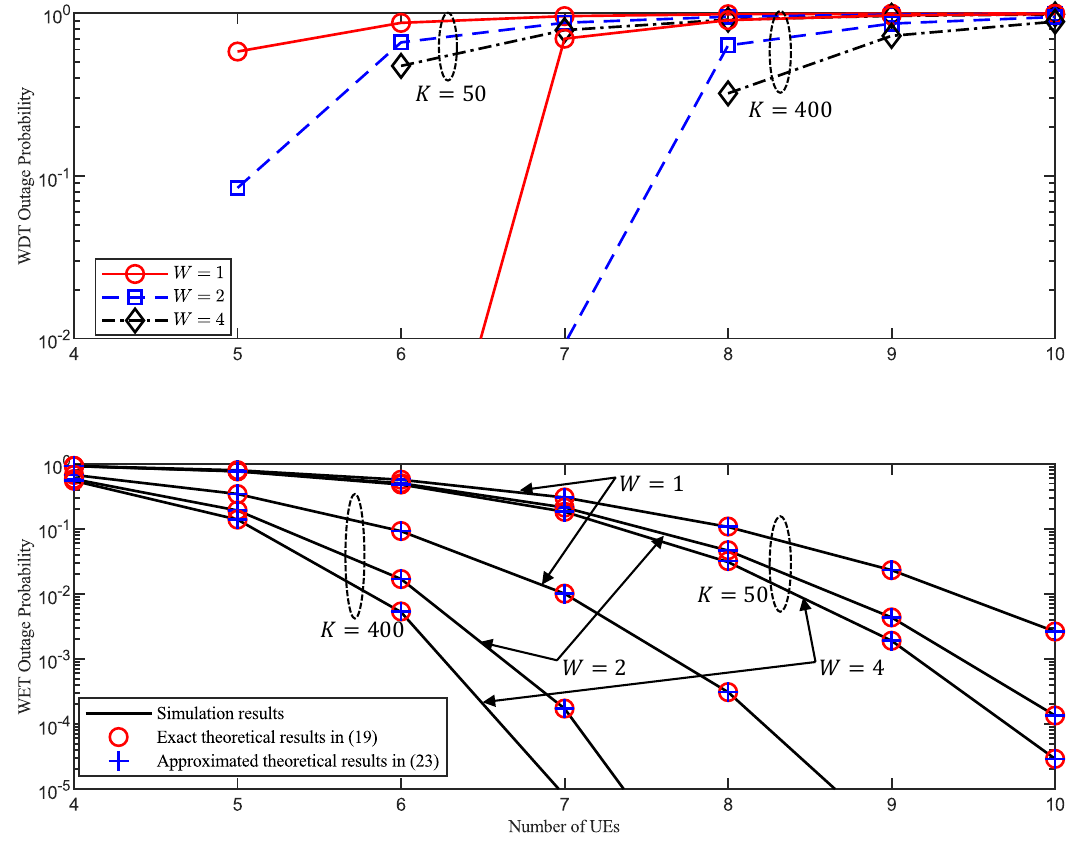}
		\caption{Outage probability versus the number of users $N$.}
		\label{Fignumofusers}
	\end{figure}
	\begin{figure}[t]
		\centering
		\includegraphics[width=2.47in]{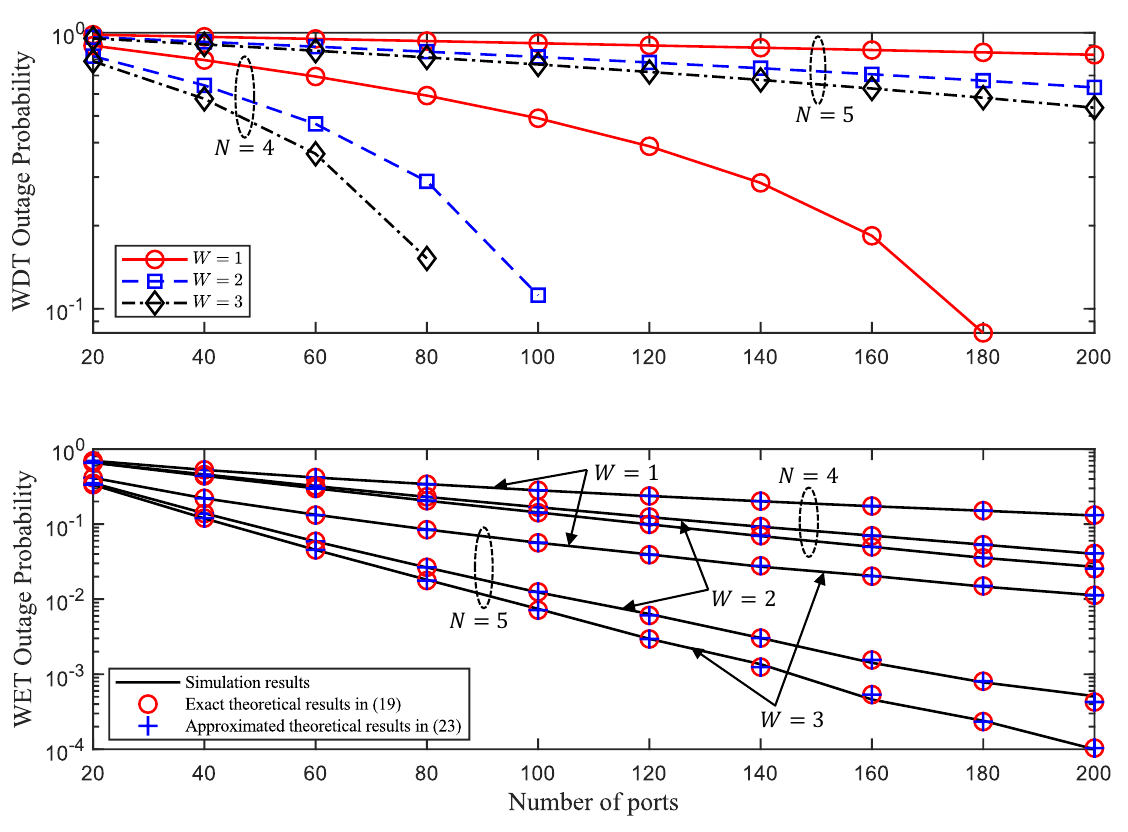}
		\caption{Outage probability versus the number of ports $K$.}
		\label{Fignumofports}
	\end{figure}
	\begin{figure}[t]
		\centering
		\includegraphics[width=2.37in]{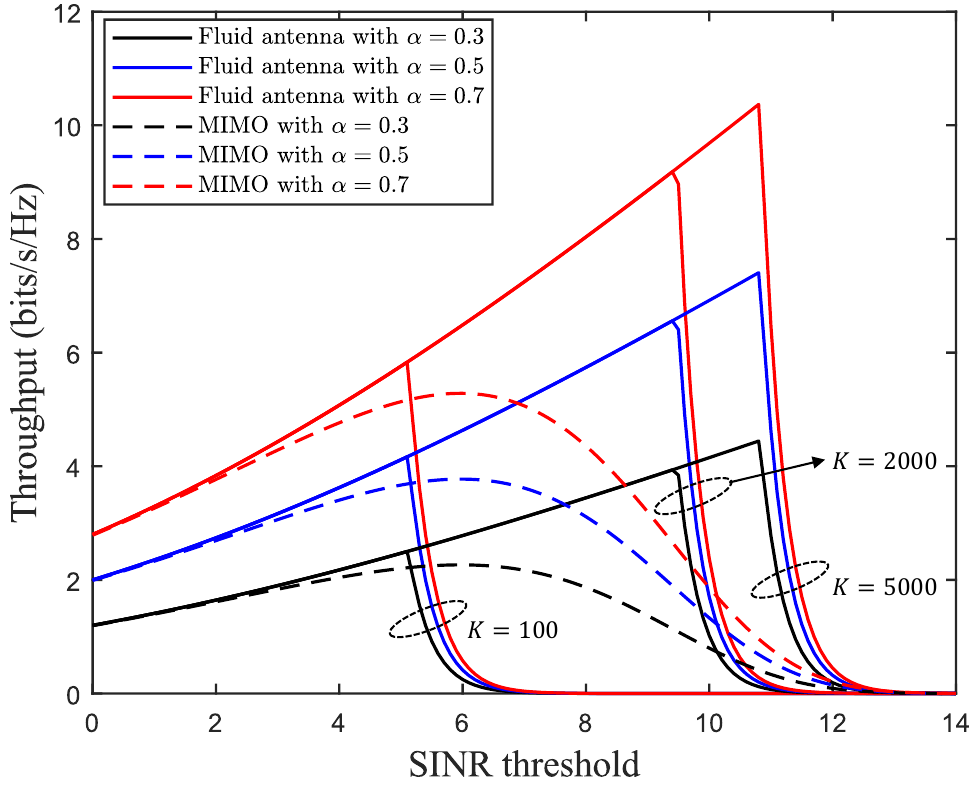}
		\caption{Reliable throughput against the SINR threshold $\gamma$.}
		\label{Figthroughput}
	\end{figure}
	\begin{figure}[t]
		\centering
		\includegraphics[width=2.42in]{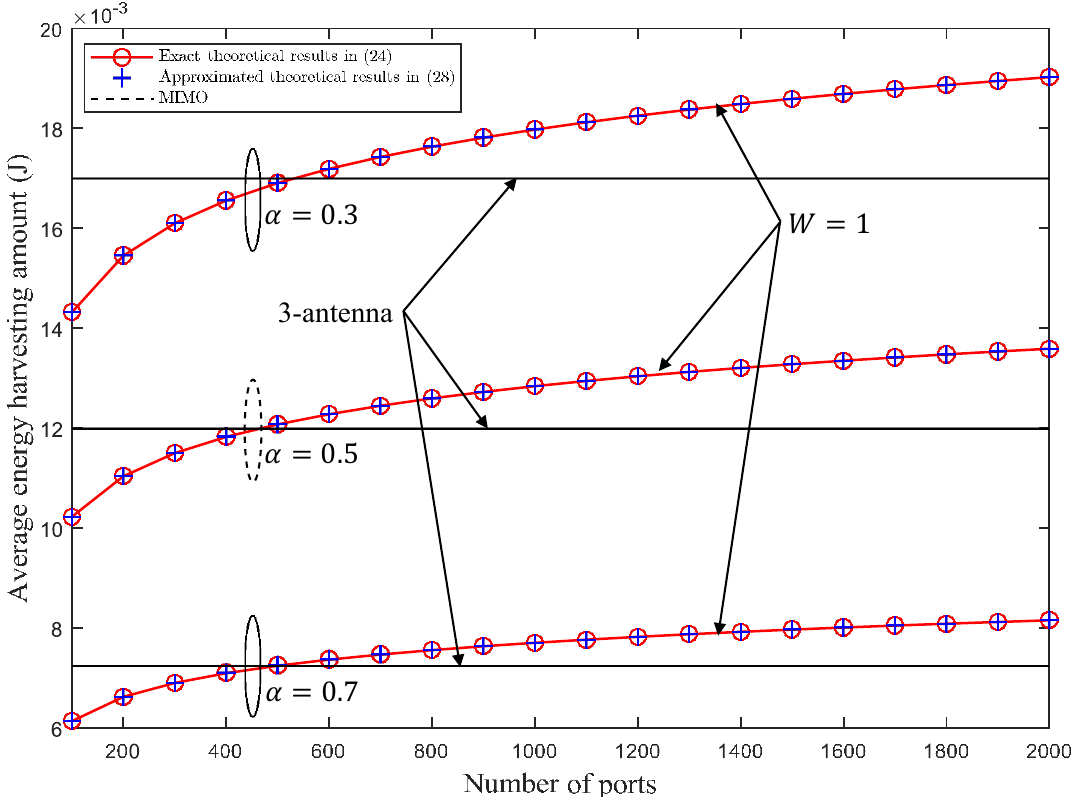}
		\caption{Aaverage energy harvesting amount against the number of ports $K$.}
		\label{FigavgE}
	\end{figure}
	In this section, the performance of the FAMA assisted IDET system is evaluated by both the theoretical analysis and Monte Carlo based simulation. It is assumed that the wireless channels of different AP-UE pairs have identical statistical characteristics. The transmission power of all the APs are set to the same as $P=1$ W, while the duration of a period is set to $T=1$ ms.
	
	Fig. \ref{Fignumofusers} depicts the WDT outage probability and the WET outage probability versus the number of UEs, $N$, with $\gamma=3.6$ dB and $Q_{th}=25$ W\footnote{Since the AWGN channel is conceived in our system,  the energy harvesting threshold $Q_{th}$ is an equivalent value by neglecting the impact of pathloss. Note that by defining the pathloss as $\Omega$, the exact energy harvesting threshold at the receiver is $Q_{th}/\Omega$. }, where different numbers of ports and fluid antenna sizes are conceived. The WDT outage probability is obtained according to the theoretical upper bound derived in \eqref{WDTP}, while the WET outage probability is obtained by both the Monte Carlo based simulation and the derived closed-form in \eqref{final} and \eqref{approp}. Observe from Fig. \ref{Fignumofusers} that the theoretical results of the WET outage probability match well with the simulation ones, which validates the accuracy of our theoretical analysis. Moreover, the WET outage probability decreases when we increase the number of UEs $N$. This is because the UE is able to receive more wireless signals from other APs and glean more energy. On the contrary, the WDT outage probability increases with $N$, since the interference from other APs becomes much serious. Furthermore, it can be observed that a larger number of ports $K$, or a larger size of fluid antenna $W$, concurrently improve the performance of both WDT and WET. This is because a larger $K$ or $W$ increases the likelihood of selecting an optimal port having a better wireless channel gain, which improves the received signal strength for either WDT or WET.  Moreover, a trade-off between the WDT and WET performance should be obtained by optimizing the number of UEs in the network.
	
	Fig. \ref{Fignumofports} depicts the WDT outage probability and the WET outage probability versus the number of ports $K$, with $\gamma=6.1$ dB and $Q_{th}=18$ W, where different numbers of UEs and fluid antenna sizes are conceived. As expected, the WDT outage probability and the WET outage probability decrease when $K$ or $W$ increases. Note that when the antenna size $W$, increases from $W=1$ to $W=2$, the outage probalities of both WDT and WET significantly decrease, which demonstrates that the antenna size is a critical influencing factor for both WDT and WET.
	
	Fig. \ref{Figthroughput} illustrates the reliable throughput by comparing the fluid antenna system and the 3-antenna MIMO system in \cite{FAsystem} with the same antenna size. Observe from Fig. \ref{Figthroughput} that the reliable throughput is concave with respect to the SINR threshold $\gamma$ for both systems. This is because a larger SINR threshold improves the transmission data rate, which results in the increase of the reliable throughput. However, when the SINR threshold is too high, the outage probability of WDT increases rapidly, which dominates the reliable throughput and lead to the further decreasement. As expected, the reliable throughput increases with the TS ratio $\alpha$. Moreover, the fluid antenna system is able to achieve a better reliable throughput  performance compared to the traditional MIMO benchmark, when the number of ports is large enough, \textit{i.e.}, $K=5000$.
	
	Fig. \ref{FigavgE} investigates the average energy harvesting amount $E_{avg}$ versus the number of ports $K$, where the antenna size is set to $W=1$ and the number of UEs is $N=4$. Observe from Fig. \ref{FigavgE} that when we increase $K$, the average energy harvesting amount also increase, since a larger $K$ enhances the probability of selecting an optimal port having a better wireless channel gain, which results in an improved received signal strength for WET. Moreover, we also compare the WET performance of the fluid antenna system with that of a $3$-antenna MIMO system in \cite{Optimal} with the same antenna size. It is observed that FAMA assisted IDET system outperforms the $3$-antenna MIMO system when $K$ is large enough, \textit{i.e.}, $K=800$. As expected, a larger TS ratio $\alpha$ results in a worse energy harvesting performance, since the WET duration is shortened. Therefore, a trade-off between the WDT and WET performance should be obtained by optimizing the TS ratio in the network.

	\section{Conclusion}
	In this paper, a FAMA assisted IDET system was studied, where a fluid antenna is equipped at each UE to mitigate the multiuser interference and enhance the strength of energy harvesting signals by dynamically switching the antenna ports. Simulation results validated the theoretical analysis and evaluated the IDET performance of the system, which demonstrated that the FAMA assisted system could achieve a better IDET performance than traditional MIMO, while the trade-off between WDT and WET should be obtained by optimizing the number of users and TS ratio of the network.

	\vspace{12pt}

\begin{thebibliography}{00}	
	    
		\bibitem{Z.Zhang} Z. Zhang et al., ``6G Wireless Networks: Vision, Requirements, Architecture, and Key Technologies," \emph{IEEE Veh. Technol. Mag.}, vol. 14, no. 3, pp. 28-41, Sept. 2019.
			
		\bibitem{W.Saad}W. Saad, M. Bennis and M. Chen, ``A Vision of 6G Wireless Systems: Applications, Trends, Technologies, and Open Research Problems," \emph{IEEE Network.}, vol. 34, no. 3, pp. 134-142, May/June 2020.
		
		\bibitem{FAsystem} K. K. Wong, A. Shojaeifard, K. -F. Tong and Y. Zhang, ``Fluid Antenna Systems," \emph{IEEE Trans. Wireless Commun.}, vol. 20, no. 3, pp. 1950-1962, March 2021.
		
		\bibitem{FAMA} K. K. Wong, and K. F. Tong, ``Fluid antenna multiple access,” \emph{IEEE Trans. Wireless Commun.}, vol. 21, no. 7, pp. 4801–4815, Jul. 2022.
		
		\bibitem{gkbykit} K. -K. Wong, D. Morales-Jimenez, K. -F. Tong and C. -B. Chae, ``Slow Fluid Antenna Multiple Access," \emph{IEEE Trans. Commun.}, vol. 71, no. 5, pp. 2831-2846, May 2023.
		
		\bibitem{Massive} K. -K. Wong, K. -F. Tong, Y. Chen and Y. Zhang, ``Fast Fluid Antenna Multiple Access Enabling Massive Connectivity," \emph{IEEE Commun.  Lett.}, vol. 27, no. 2, pp. 711-715, Feb. 2023
		
		\bibitem{yang} H. Yang, K. -K. Wong, K. -F. Tong, Y. Zhang and C. -B. Chae, ``Performance Analysis of Slow Fluid Antenna Multiple Access in Noisy Channels Using Gauss-Laguerre and Gauss-Hermite Quadratures," \emph{IEEE Commun. Lett.}, vol. 27, no. 7, pp. 1734-1738, July 2023.
		
		\bibitem{Oppo} K. -K. Wong, K. -F. Tong, Y. Chen, Y. Zhang and C. -B. Chae, ``Opportunistic Fluid Antenna Multiple Access," \emph{IEEE Trans. Wireless Commun.}, vol. 22, no. 11, pp. 7819-7833, Nov. 2023
		
		\bibitem{wu1} Y. Wu, S. Feng and C. Dong, ``Energy Constrained Data Collection in Multi-UAV-Assisted IoT,"  \emph{2023 IEEE 97th Vehicular Technology Conference (VTC2023-Spring)}, Florence, Italy, 2023, pp. 1-7.
		
		\bibitem{wu2}	J. Zhang, Y. Wu, G. Min and K. Li, ``Neural Network-Based Game Theory for Scalable Offloading in Vehicular Edge Computing: A Transfer Learning Approach," early access in \emph{IEEE Trans. Intell. Transp. Syst.}
		
		\bibitem{Rzhang} R. Zhang and C. K. Ho, ``MIMO Broadcasting for Simultaneous Wireless Information and Power Transfer," \emph{IEEE Trans. Wireless Commun.}, vol. 12, no. 5, pp. 1989-2001, May 2013.
	
		\bibitem{Optimal} Z. Zong, H. Feng, F. R. Yu, N. Zhao, T. Yang and B. Hu, ``Optimal Transceiver Design for SWIPT in  K-User MIMO Interference Channels," \emph{IEEE Trans. Wireless Commun.}, vol. 15, no. 1, pp. 430-445, Jan. 2016.
		
		\bibitem{hujie} J. Hu, K. Yang, G. Wen and L. Hanzo, ``Integrated Data and Energy Communication Network: A Comprehensive Survey," \emph{IEEE Commun. Surveys \& Tuts.}, vol. 20, no. 4, pp. 3169-3219, Fourthquarter 2018.
		
		\bibitem{mu}K. K. Wong, K. F. Tong, Y. Chen and Y. Zhang, ``Closed-form expressions for spatial correlation parameters for performance analysis of fluid antenna systems,” \emph{IET Elect. Lett.}, vol. 58, no. 11, pp. 454–457, Apr. 2022.
		
		\bibitem{Relaying}A. A. Nasir, X. Zhou, S. Durrani and R. A. Kennedy, ``Relaying Protocols for Wireless Energy Harvesting and Information Processing," \emph{IEEE Trans. Wireless Commun.}, vol. 12, no. 7, pp. 3622-3636, July 2013.
		
		\bibitem{HGV} Simon, Marvin Kenneth. \emph{Probability distributions involving Gaussian random variables: A handbook for engineers and scientists}. Boston‐Dordrecht‐London: Kluwer Academic Publishers, 2002.
		
		\bibitem{gaussian} M. Abramowitz and I. A. Stegun, \emph{Handbook of Mathematical Functions with Formulas, Graphs, and Mathematical Tables}, Dover Publications Inc.; New edition, Sixth Printing, Nov. 1967, with correlations.	
		
		
	\end{thebibliography}
\end{document}